\documentclass[preprint,12pt]{elsarticle}

\usepackage{amssymb}

\journal{Iranian Journal of Science}
\usepackage{amscd, amssymb}
\usepackage[mathscr]{eucal}
\usepackage{multirow}
\usepackage{dcolumn}
\newcolumntype{+}{D{/}{\mbox{--}}{4}}
\newcolumntype{,}{D{,}{,}{2}}
\newtheorem{thm}{Theorem}[section]
\newtheorem{lem}[thm]{Lemma}
\newtheorem{ex}[thm]{Example}
\newtheorem{defn}[thm]{Definition}

\newtheorem{alg}[thm]{Algorithm}
\newtheorem{rem}[thm]{Remark}

\newcommand{\eep}{\hfill $\square$}
\newcommand{\pf}{\noindent {\bf Proof. \ }}
\usepackage{algorithm}
\usepackage{algorithmic}
\usepackage{epsfig}
\begin{document}

\begin{frontmatter}



\title{An Efficient Algorithm for Counting Cycles in QC and APM LDPC Codes}

\author{Mohammad Gholami\corref{cor1}\fnref{label1,label2}}
\ead{gholami-m@sku.ac.ir}
 \fntext[label1]{Department of Mathematics, Shahrekord University, P.O. Box: 88186-34141, Shahrekord, Iran}
 \fntext[label2]{Institute for Research in Fundamental Sciences (IPM), P.O. Box: 19395-5746, Tehran, Iran,}
 \cortext[cor1]{Corresponding author}
\author{Zahra Gholami\fnref{label1}}
\ead{zghbaba123@gmail.com}
%



\begin{abstract}
In this paper, a new method is given for counting cycles in the Tanner graph of a (Type-I) quasi-cyclic (QC) low-density parity-check (LDPC) code
which the complexity mainly is dependent on the base matrix, independent from the CPM-size of the constructed code.
Interestingly, for large CPM-sizes, in comparison of the existing methods, this algorithm is the first approach which efficiently counts the cycles in the Tanner graphs of QC-LDPC codes. In fact, the algorithm recursively counts the cycles in the parity-check matrix column-by-column by finding all non-isomorph tailless backtrackless closed (TBC) walks in the base graph and enumerating theoretically their corresponding cycles in the same equivalent class. Moreover, this approach can be modified in few steps to find  the cycle distributions of a class of LDPC codes based on Affine permutation matrices (APM-LDPC codes).
Interestingly, unlike the existing methods which count the cycles up to $2g-2$, where $g$ is the girth, the proposed algorithm can be used to enumerate the cycles of arbitrary length in the Tanner graph. Moreover, the proposed cycle searching algorithm improves upon various previously
known methods, in terms of computational complexity and memory requirements.
\end{abstract}



\begin{keyword}


QC-LDPC codes\sep Tanner graph\sep girth\sep counting cycles\sep backtrackless closed walks.
\end{keyword}
\end{frontmatter}






\section{Introduction}
Low-density parity-check (LDPC) codes, first discovered by Gallager \cite{gal}, were rediscovered about 30 years later to be Shannon limit-approaching codes over additive white Gaussian noise (AWGN) channels \cite{9},\cite{14}. It is established that the performance of LDPC codes under iterative decoding depends upon by certain combinatorial structures such as multiplicities and distribution of short cycles in the Tanner graph. When there are short cycles in the Tanner graph, the belief-propagation algorithm (BPA) does not converge to maximum likelihood performance \cite{8}. Because, a message delivered by a node along a cycle will propagate back to the node itself after several iterations causing decreasing of independence in the messages sent afterward.

Large girth, however, is not sufficient  to ensure a good graphical model. The performance of two LDPCs with the same girth, but a different
number of short cycles, can be significantly different. The regularity or non-regularity of the cycle structure of a graph (i.e., how a randomness effect of the graph appears) also affects graphical code model quality. For example, the introduction of irregularity to LDPC designs is known to improve performance \cite{12}. In summary, good graphical models of codes imply  to have large girth, small number of short cycles, or some cycle structures which are not
overly regular. Based on the connection between the performance of a code and the properties of its associated graphical model, characterizing
the cycle structure of a graphical model is of great interest. The difficulty in enumerating  and counting cycles and paths in arbitrary graphs  may
prevent an efficient search of good LDPC codes with small short cycles. To solve the problem, this paper presents a new algorithm of counting short cycles  by analyzing the shapes of the cycles of Tanner graph in base matrix for designing good LDPC codes which is
less complex than the existing algorithms.

Unlike the existing algorithms based on matrix multiplication \cite{7}, \cite{14} which cause to high computational complexity and time consuming, the proposed method has a lower complexity for counting short cycles. In the proposed method, first all of the chains associated to TBC walks in the protograph of a QC-LDPC code are classified by a non-isomorphic relation, then each chain corresponds to some cycles which are enumerated by the period of some sets (in fact, these sets are the row indices of the nonzero elements in a block which have the same topological shapes when traversing the walk by starting from each element of the set).
This method can be used effectively to evaluate the performance of LDPC codes according to their short
circle distributions.

Several methods have been investigated the parity-check matrices such that the associated TGs are free of short cycles \cite{9},~\cite{14}. However, LDPC codes are often designed without explicit constraints on the girth. In \cite{114}, a method for counting cycles of length less than 8 is presented, however, this method is complex for longer cycles and addresses the restricted problem of counting the cycles of length $g$, $g+2$ and $g+4$ in bipartite
graphs with girth $g$. In \cite{newcounting}, a novel theoretical method is proposed to evaluate the number of closed paths of different lengths in an
all-one base matrix up to closed paths of length 10. The algorithm in \cite{7}, \cite{11}, which is capable of counting short cycles
$g, g+1, \cdots, 2g-1$ in a general graph, and short cycles of length $g, g+2, \cdots , 2g-2$ in the Tanner graph. The algorithm \cite{11}, is based on performing integer additions and subtractions in the nodes of the graph and passing extrinsic messages to adjacent nodes. The complexity of the method \cite{11} is $O(g|E|^2)$, where $|E|$ is the number of edges in the graph. The proposed method in \cite{Brevik} is based on the relationship between the number of short cycles in the graph and the eigenvalues of the directed edge-adjacency matrix of the graph. In order to find the eigenvalues of the directed edge matrix of
the graph, one needs to find the eigenvalues of $N$ matrices, each of size
$|E_b| \times |E_b|$. This reduces the complexity from $O(N^3|E_b|^3)$ to $O(N|E_b|^3)$, which $|E_b|$ is the number of directed  edges in matrix of the base graph. Compared to the complexity $O(g|E|^2)=O(gN^2|E_b|^2)$ of the algorithm in \cite{11}, the proposed algorithm
is less complex if $gN$ grows faster than $|E_b|$. This would be the case for protograph codes with small base graph and large
lifting degree. In terms of memory requirements, for example, for a regular
LDPC code of variable node degree $d_u$, the proposed method
needs $N|E_b|^2$ memory location eigenvalues of $N$ matrices. The algorithm of
\cite{11}, on the other hand, needs $2d_u|E|$ locations, which would
be less than that of the proposed method if $|U_b|>2$.

For fixed values of $k\leq 13$, there are explicit formula \cite{1}, \cite{2} and \cite{3}, expressing the number of cycles of length $k$ through adjacency matrix graph. The computational complexity of these formulas for $k\leq 7$ is of the same order as the multiplication of $n \times n$ matrices, and for $k \geq 8$, it is the amount of $O(n^{[k/ 2]}\log n)$, where $n$ is the number of vertices.  
It is known that the complexity of counting cycles of length $k$ in
 arbitrary graphs inevitably increases with $k$ \cite{6}.
 A number of works devoted  counting short cycles in bipartite graphs LDPC-code is characterized by low density and the value of the girth $g \geq 6$. The feature most the methods discussed below is that by limiting the length of the cycle (relative
 to  girth) is achieved while the $O(n^4)$.  Thus, the authors of \cite{7} (halford) presented an algorithm
 for girth $g$ bipartite graph, and count the number of cycles of length $g, g +2$ and $g +4$
 with the same order of complexity that multiplication $n \times n$ matrices (for fixed $g$). They showed that in addition to the girth,
the number and statistics of short cycles are also important
performance metrics of the code. The complexity of
their method is $O(gn^3)$, where $n$ is the size of the larger set
between the two node partitions. 

In~\cite{20},~\cite{myung}, some affine permutation matrices (APM) are used to generate a class of
LDPC codes, called APM-LDPC codes, which are not QC in general. Unlike Type-I conventional QC-LDPC codes,
the constructed $(J,L)$ APM-LDPC codes with the
$J\times L$ all-one base matrix can achieve minimum distance greater than
$(J+1)!$ and girth larger than 12.
 Moreover, the lengths of the constructed
APM-LDPC codes, in some cases, are smaller than the best known lengths
reported for QC-LDPC codes with the same base matrices. As an advantage, the
constructed APM-LDPC codes are flexible in lengths and rates. In
some cases the lengths of the constructed codes are smaller than the
best known lengths reported for the lengths of QC-LDPC codes
in~\cite{boch},~\cite{karimi},~\cite{fos},~\cite{sulivan},~\cite{samad},~\cite{girth18}. Another significant advantage of the constructed APM-LDPC codes is that they have remarkably fewer cycle multiplicities compared to QC-LDPC codes with the same base matrices and the same lengths.
Simulation results show that the constructed APM-LDPC codes with lower girth
outperform QC-LDPC codes with larger girth.

This paper is organized as follows.
In Section \ref{sec1}, first, some preliminaries and notations useful for the next sections of the paper, are provided. Then, cycles of QC-LDPC codes and APM-LDPC codes are investigated in Section \ref{sec2} by some modular equations and allowable and non-allowable chains are defined to estimate the number of cycles in the Tanner graph of a QC (APM) LDPC code. Finally, for a given binary matrix $B$, an algorithm for counting the cycles of a QC (APM) LDPC codes is introduced in Section \ref{sec4} and then, the complexity of this algorithm is investigated.

\section{Preliminaries and Definitions}
\label{sec1}
An undirected {\it Graph} $G=(V,E)$ is defined as a set of nodes $V$ and a set of edges $E$, where $E$ is some subset of
the pairs $\{\{u, v\} : u, v\in V, u \neq v\}$. A {\it walk} of length $k$ in $G$ is a sequence of nodes $v_1, v_2, \cdots , v_{k+1}$ in $V$ such that $\{v_i, v_{i+1}\}\in E$ for all $i\in \{1, \cdots , k\}$. Equivalently, a walk of length $k$ can be described
by the corresponding sequence of $k$ edges. 
A walk is {\it closed} if the two end nodes are identical, i.e.,
$v_1=v_{k+1}$ in the previous description. 
A closed walk is {\it backtrackless} if $e_{i_s}
\neq e_{i_s+1}$ for each $1\le s\le k$ and it is {\it tailless} if $e_{i_1}
\neq e_{i_k}$. 
 Let $v,k$ be some positive integers and $V=\{1,2,\ldots,v\}$. By a $(v,k)$ block-design, we mean a list of $k$ subsets $B_i$, $1\le i\le k$, of $V$,  denoted by ${\cal B}=[B_1,B_2,\ldots,B_k]$. In this definition, $B_i$, $1\le i\le k$, are called the blocks and the term {\em list} is used to allow the {\em repetition} and the {\em ordering} of the blocks. To an LDPC code with the parity-check matrix $H$, a bipartite graph, called {\it Tanner graph} TG$(H)$, is associated
which collects variable nodes and check nodes corresponding to the columns and rows of $H$, respectively, and each edge connects a check node to a bit node if nonzero entry exists in the intersection of the corresponding row and column of $H$. The girth of a code with a given parity-check matrix $H$, denoted by $g(H)$, is the length of a shortest cycle in
TG$(H)$ which is always an even number.

{\it Protograph codes}~\cite{throp} are a class of structured LDPC codes, constructed from a bipartite graph with relatively small number of variable nodes and check nodes, called a {\it protograph}. In the construction, the first step is to choose a
protograph with a near capacity decoding threshold as a building block
and then to make copies of the chosen protograph and permute the edges of copies according
to certain rules to connect them into a Tanner graph of larger
size. The parity-check matrix of a protograph code can be obtained from the incidence matrix of protograph with the replacement of each 1 and 0 by some $m\times m$ permutation and zero matrices, respectively. By considering such permutations as circulant permutation matrices (CPM) or affine permutation matrices (APM), two classes of protograph codes, called {\it (Type I) QC-LDPC codes} and {\it APM-LDPC codes}, respectively, can be defined. For some integers $m$, $s$ and $a$, satisfying in $0\le s<m$, $1\le a<m$ and $\gcd(a,m)=1$, by the APM $I_m^{s,a}$ with slope $s$ and shift $a$, briefly  $I^{s,a}$ when $m$ is known, we mean a $m\times m$ binary matrix $(e_{i,j})_{0\le i,j<m}$ in which $e_{i,j}=1$ if and only if
$i=aj+s\bmod m$. In fact, $I^{s,a}$ is the $m\times m$ binary permutation matrix for which the only non-zero
element in the first column occurs in position $s$, and each other
column is shifted down by  $a$ positions, regard to the previous
column. In particular, if $a=1$, $I_m^{s,a}$ is denoted by $I_m^s$ which is called a CPM of size $m$ and slope $s$. Hence, APM LDPC codes can be considered as a generalization of QC-LDPC codes.

Based on the evidence obtained that connecting the performance of a code and the properties of its associated graphical model, characterizing
the cycle structure of a graphical model is of great interest. The difficulty in enumerating  and counting cycles and paths in arbitrary graphs  may
prevent an efficient search of good LDPC codes with small short cycles. To solve the problem, we first present
a new recursive algorithm for counting short cycles in QC LDPC codes based on analyzing the TBC walks
in protograph having less complexity rather than the other existing methods. Then, we use a modified version of this algorithm to count the cycles in APM-LDPC codes.
\section{\large Cycles in QC $\&$ APM LDPC codes\label{sec2}}
As QC-LDPC codes can be embraced in the class of APM LDPC codes and for simplicity of notations,  we first set up the notations and terminologies for the case of APM-LDPC codes, then we apply them for QC-LDPC codes.
For some positive integers $v$, $k$, $v<k$, let $B$ be a $v\times k$ binary matrix and ${\cal B}=[B_1,B_2,\ldots,B_k]$ be the corresponding $(v,k)$ block-design with blocks $B_i\subseteq V=\{1,2,\ldots,v\}$, where each $B_i$ is the row-indices
of non-zero elements of the $i$th column of $B$. For positive integer $m$, by a $(m,{\cal B})$-slope vector and a $(m,{\cal B})$
shift vector, we mean two finite sequences $S=(s_{i,j})_{1\le i\le v, j\in B_i}$ and $A=(a_{i,j})_{1\le i\le v, j\in B_i}$, respectively,
such that each $s_{i,j}$ belongs to ${\Bbb Z}_m$ and $a_{i,j}$ belongs to ${\Bbb Z}^*_m$, where ${\Bbb Z}_m$ is the ring of integers modulo $m$ and ${\Bbb Z}^*_m=\{a\in{\Bbb Z}:\gcd(a,m)=1\}$. Now, for given $(m,{\cal B})$-slope vector $S$ and $(m,{\cal B})$-shift vector $A$, let ${\cal H}_{m,{\cal B},S,A}$ be the $vm\times km$ parity-check matrix of an APM-LDPC code with APM size $m$
obtained by replacing each zero and $(i,j)$ non-zero element of $B$
by the $m\times m$ zero matrix and $I^{s_{i,j},a_{i,j}}$, respectively. For the case of QC-LDPC codes, i.e. when $A$ is a fully one matrix, we just use ${\cal H}_{m,{\cal B},S}$ to denote the corresponding parity-check matrix. Now, the following theorem is very useful to verify the cycles in the Tanner graph of each APM-LDPC code.
\begin{thm}\rm(\cite{20})
Each $2l-$cycle in TG$({\cal H}_{m,{\cal B},S,A})$ corresponds to a finite chain
$(i_{0},$ ${j_0},$ $i_{1},$ ${j_1},$ $\ldots,$ $i_{{l-1}},$ ${j_{l-1}})$, $i_{l}=i_{0},$
such that for each $1\le k\le l$, $\{i_{k-1}, i_{k}\}\subseteq B_{j_{k-1}}$, $i_{k-1}\ne i_{k}$, $j_{k-1}\ne j_{k}$, and for $A=\sum_{k=0}^{l-1}(p_ks_{i_k,j_k}-p_{k+1}s_{i_{k+1,j_k}})$, where
$p_h=\prod_{k=h}^{l-1}a_{i_{k+1},j_k}a_{i_{k},j_k}^{-1}\bmod m$, $0\le h\le l-1$, one of the following relations holds:
\begin{enumerate}
\item $p_0=1$ and $A=0$.
\item $\gcd(p_0-1,m)|A$.
\end{enumerate}
\label{thm-apm}
\end{thm}
Especially, for QC-LDPC codes, Theorem~\ref{thm-apm} can be simplified as follows.
\begin{thm}\rm(\cite{fos})
Each $2l-$cycle in TG$({\cal H}_{m,{\cal B},S})$ corresponds to a $2l$-cycle chain $(i_{0},$ ${j_0},$ $i_{1},$ ${j_1},$ $\ldots,$ $i_{{l-1}},$ ${j_{l-1}})$ in which $\sum_{k=0}^{l-1}(s_{i_k,j_k}-s_{i_{k+1,j_k}})=0\pmod m$.
\label{thm-qc}
\end{thm}
Hereinafter, each finite chain $(i_{0},$ ${j_0},$ $i_{1},$ ${j_1},$ $\ldots,$ $i_{{l-1}},$ ${j_{l-1}})$ satisfying in Theorem~\ref{thm-apm} for APM-LDPC codes (or Theorem~\ref{thm-qc} for QC LDPC codes) is called a {\it $2l$-cycle chain}.
\begin{figure}
\begin{center}
\includegraphics[scale=1]{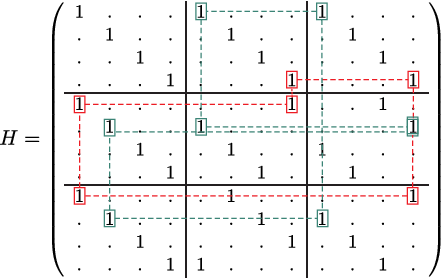}
\end{center}
\caption{\small An unallowable 14-cycle sequence.} \label{fig2}
\end{figure}

The converse of Theorem~\ref{thm-apm} is not true in general. In other words, each $2l-$cycle chain may be induces a cycle in the Tanner graph with length less than $2l$. In this case, this cycle chain contains a $2k$-cycle chain, for some $k<l$ and some middle points in the cycle return on their owns. This situation describes a {\it non-allowable} cycle chain, otherwise, i.e. if the cycle chain doesn't contain an smaller cycle chain, we say that the cycle chain is {\it allowable}. For example, Fig.~\ref{fig2} shows the cycle chain $(2,2,0,1,1,2,0,1,1,0,2,2,1,0)$ which is non-allowable, because it contains an smaller cycle chain $(2,2,0,1,1,0,2,2)$. For non-allowable cycle chains, it is noticed that the elements of the proper sub-chain are not essentially successive in the parent cycle chain.

In continue, we give a necessary and sufficient condition for a cycle-chain in a APM LDPC code to be allowable.
Let $L=(i_{0},{j_0},i_{1},{j_1},\ldots,i_{{l-1}},{j_{l-1}})$ be a $2l$-cycle chain in a APM LDPC code with $(m,{\cal B})$-slope vector $S=(s_{i,j})_{1\le i\le v, j\in B_i}$ and $(m,{\cal B})$ shift vector $A=(a_{i,j})_{1\le i\le v, j\in B_i}$ where ${\cal B}=[B_1,B_2,\ldots,B_k]$ is a $(v,k)$ block-design for some $v,k$. Now, for a given $0\le r\le m-1$, define the sequences $(\delta^{\rightarrow,t})_{t=0}^{l-1}$ and $(\delta^{\uparrow,t})_{t=0}^{l-1}$ recursively, as follows:
$$\left\{\begin{array}{ll}
\delta^{\rightarrow,0}=r,& \delta^{\rightarrow,t+1}=a_{i_{t+1},j_t}a_{i_{t},j_t}^{-1}(\delta^{\rightarrow,t}-s_{i_{t},j_t})+s_{i_{t+1},j_t}\bmod m, \\ \\ \delta^{\uparrow,0}=a_{i_0,j_0}^{-1}(r-s_{i_0,j_0})\bmod m,& \delta^{\uparrow,t+1}=a_{i_{t+1},j_{t+1}}^{-1}(a_{i_{t+1},j_t}\delta^{\uparrow,t}+s_{i_{t+1},j_t}-s_{i_{t+1},j_{t+1}})\bmod m
\end{array}\right.$$

\begin{lem}\rm
$L$ is allowable if and only if for each $p,q,r$, $0\le p<q\le l-1$, $0\le r\le m-1$, we have $\delta^{\rightarrow,q}\ne\delta^{\rightarrow,p}$ when $i_p=i_q$, and $\delta^{\uparrow,q}\ne\delta^{\uparrow,p}$ when $j_{p}=j_q$.

\pf Corresponding to the cycle chain $L=(i_{0},{j_0},i_{1},{j_1},\ldots,i_{{l-1}},{j_{l-1}})$, let ${\cal P}_r=v_{1}v_2\ldots v_{2l}v_{2l+1}$, $v_1=v_{2l+1}$, be the closed path starting from the point $v_1$ having row-index $r$ in the block $(i_0,j_0)$ of the parity-check matrix $H$.  Without loss of generality, let $v_2$ be the point having the same column-index with $v_1$, therefore each vertex $v_k$, $1\le k\le 2l$, with the odd or even index $k$ belongs to the block $(i_p,j_p)$ or $(i_{p},j_{p-1})$, respectively, where $p=\lfloor\frac{k}{2}\rfloor$. Clearly, the column-index of $v_1$ is $a_{i_0,j_0}^{-1}(r-s_{i_0,j_0})\bmod m$. Moreover, for each $1\le k\le 2l$, it can be seen easily that $\delta^{\rightarrow,k}$ and $\delta^{\uparrow,k}$ are the row and column indices of the point $v_{k}$, respectively. Now, if $L$ in not allowable, then for some $0\le r\le m-1$ and $1\le k+1<k'\le 2l$, two middle (not successive) points $v_{k}$ and $v_{k'}$  belong to the same column-block of $H$, whereas they have the same column-indices, i.e. $\delta^{\uparrow,p}=\delta^{\uparrow,q}$ and $j_p=j_q$, where $p=\lfloor\frac{k}{2}\rfloor<q=\lfloor\frac{k'}{2}\rfloor$, or $v_{k}$ and $v_{k'}$ belong to the same row-block of $H$, while their row-indices are the same, i.e. $\delta^{\rightarrow,p}=\delta^{\rightarrow,q}$ and $i_p=i_q$. Now, the proof is completed.\eep
\label{lem-allow}
\end{lem}


For QC-LDPC codes, allowability of cycle-chains in Lemma~\ref{lem-allow} can be simplified as follows.
\begin{lem}\rm
Let $L=(i_{0},{j_0},i_{1},{j_1},\ldots,i_{{l-1}},{j_{l-1}})$ be a $2l$-cycle chain in a QC LDPC code with $(m,{\cal B})$-slope vector $S=(s_{i,j})_{1\le i\le v, j\in B_i}$, where ${\cal B}=[B_1,B_2,\ldots,B_k]$ is a $(v,k)$ block-design for some $v,k$. Then, $L$ is allowable if and only if for each $p,q$, $0\le p<q\le l-1$, if we define $\Delta_{p,q}^{\rightarrow}=\sum_{k=p}^{q}(s_{i_{k+1},j_k}-s_{i_k,j_k})\bmod m$ and $\Delta_{p,q}^{\uparrow}=\sum_{k=p}^{q}(s_{i_{k+1},j_{k+1}}-s_{i_{k+1},j_k})\bmod m$, then for each $0\le p<q<l-1$, we have $\Delta_{p,q}^{\rightarrow}\ne 0$ when $i_{p}=i_q$, and $\Delta_{p,q}^{\uparrow}\ne 0$, when $j_{p}=j_q$.
\label{lem-allow-qc}
\end{lem}
\begin{figure}
\begin{center}
\includegraphics[scale=.7]{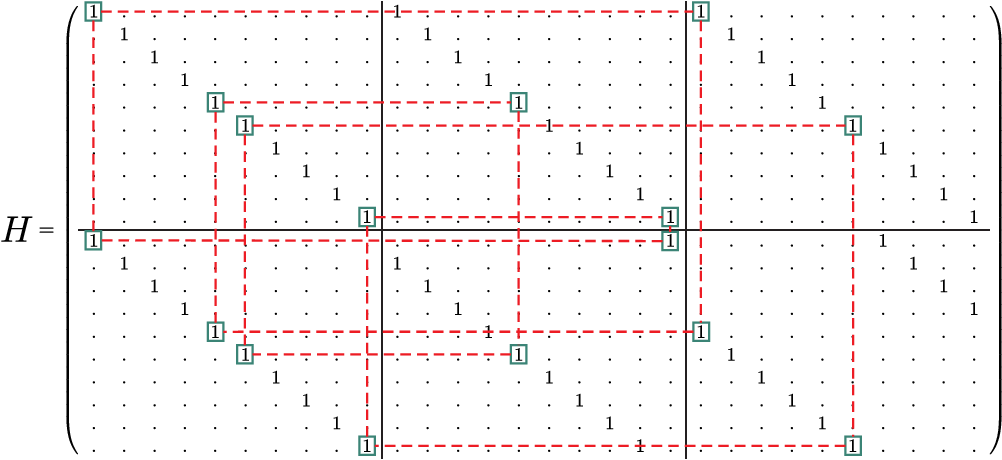}
\end{center}
\caption{\small An allowable $16-$cycle chain $(1,2,0,0,1,1,0,0,1,2,0,0,1,1,0,0)$ in $H$.} \label{f2}
\end{figure}
\begin{ex}\rm
Let $H$ be the parity-check matrix of a QC LDPC code shown in Fig.~\ref{f2} with $(10,{\cal B})$-slope vector $(s_{0,0},s_{1,0},s_{0,1},s_{1,1},s_{0,2},s_{1,2})=(0,0,0,1,0,4),$ where ${\cal B}=[\{1,2\},\{1,2\},\{1,2\}]$. The dash lines indicates the $16-$cycle chain $(1,2,0,0,1,1,0,0,1,2,0,0,1,1,0,0)$ in $H$ which corresponds to the relation
$(s_{1,2}-s_{0,2})+(s_{0,0}-s_{1,0})+(s_{1,1}-s_{0,1})+(s_{0,0}-s_{1,0})+(s_{1,2}-s_{0,2})+(s_{0,0}-s_{1,0})+(s_{1,1}-s_{0,1})+(s_{0,0}-s_{1,0})=10=0\pmod{10}$.
\label{ex2}
\end{ex}

\begin{figure}[h]
\begin{center}
\includegraphics[scale=0.9]{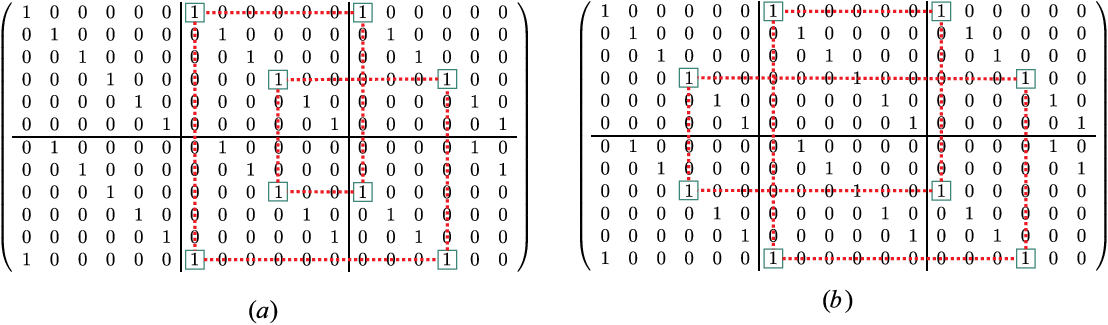}
\end{center}
\caption{\small Two allowable 8-cycles in the parity-check matrix of Example~\ref{ex1}.} \label{fig1}
\end{figure}

Hereinafter, we just consider allowable cycle chains, unless stated otherwise. \\
By Theorem~\ref{thm-qc}, cycles in the Tanner graph of a QC-LDPC code can be enumerated from cycle-chains. For this purpose, it is enough to count the whole number of cycles in the Tanner graph which corresponds to a given cycle-chain. First, we define the following relation on the set of cycle-chains.
\begin{defn}\rm
Two $2l$-cycle chains $L_1$ and $L_2$ are called {\it isomorph} if and only if there exist some $k\in\Bbb{Z}$, such that $\pi^{2k}(L_1)=L_2$, where $\pi$ is the cyclic permutation $(1,2,\ldots,2l)$ and for each permutation $\sigma\in S_{k}$,  $\sigma(i_{1},i_{2},\ldots,i_{k})$ is defined as $(i_{\sigma(1)},i_{\sigma(2)},\ldots,i_{\sigma(k)})$. In fact, each two cycle chains are isomorph if and only if they traverse the same blocks of the parity-check matrix by reading the chains by starting from different points in directions left-to-right or right-to-left. For example, in Fig.~\ref{fig1}, two 12-cycle chains $(1,2,0,0,1,2,0,1,1,0,0,1)$ and $(1,2,0,1,1,0,0,1,1,2,0,0)$ are isomorphic.
\label{def-isomorph}
\end{defn}
Clearly, the isomorph relation in Definition~\ref{def-isomorph} is an equivalence relation, which defines isomorph classes. For counting the cycles, we just consider non-isomorph cycle-sequences, which are not included in the same class.
\begin{defn}\rm
Let $(i_0,j_0,i_1,j_1,\ldots,i_{l-1},j_{l-1})$ be a $2l$-cycle sequence. For each $0\le q\le l-1$, $p=q$ or $q+1$, define $A(i_{p},j_q)$ to be the set of all values $\sum_{k=0}^{t}(s_{i_k,j_k}-s_{i_{k+1,j_k}})\bmod m$, in which the index $t$ is a nonnegative integer less than $l$ satisfied in $(i_t,j_t)=(i_p,j_q)$. In fact, $A(i_{p},j_q)$ is the row-index set of all points of the cycle when it pass from the $(i_p,j_q)$ block of the parity-check matrix. For example, in Fig.~\ref{f2}, $A(0,0)=A(1,0)=\{0,4,5,9\}$, $A(0,1)=A(1,2)=\{4,9\}$ and $A(0,2)=A(1,1)=\{0,5\}$.
\end{defn}
\begin{defn}\rm
For positive integer $m$, the period of each subset $A\subseteq \Bbb{Z}_m$, denoted by $p_m(A)$, is defined as the smallest positive number $T$ with $A+T=A$, where $A+T:=\{a+T\pmod m:a\in A\}$. For example, $\{0,2,4\}$ is a subset of $\Bbb{Z}_6$ of period 2.
\end{defn}
\begin{lem}\rm
\label{lem1}
If $A\subseteq \Bbb{Z}_m$ is of period $p_m(A)$, then $A$ can be written as the union of disjoint sets $<a_i>$, for some $a_1,\ldots,a_k\in A$, where $<a>=\{a+ip_m(A)\bmod m:1\le i\le t\}$ in which $t$ is the smallest non-negative integer satisfying $tp_m(A)=0\bmod m$.

\pf For each $a\in A$, we have $a+p_m(A)\pmod m\in A$, because $A=A+p_m(A)$, so $<a>\subseteq A$. On the other hand, $a\in <a>$, so $A=\cup_{a\in A}<a>$. Now, for each two elements $a,b\in A$, we have $<a>=<b>$ or $<a>\cap <b>=\emptyset$. Because, if $c\in<a>\cap<b>$, then $c=a+i_1p_m(A)=b+i_2p_m(A)$, for some $1\le i_1,i_2\le t$. Hence, $a=b+(i_2-i_1)p_m(A)\in <b>$ or equivalently $<a>\subseteq<b>$. Similarly, we have $<b>\subseteq<a>$, so $<a>=<b>$. Finally, $A$ can be written as union of disjoint sets $<a_i>$, for some $a_1,\ldots,a_k\in A$.\eep
\end{lem}
It is noticed that the elements $a_i$, $1\le i\le k$ in Lemma~\ref{lem1} are not necessarily unique. For example, for $m=6$ and $A=\{0,2,4\}$, we have $A=<0>=<2>=<4>$.
Now, let us mention two important consequences of the above lemma.
\begin{rem}\rm
For each $A\subseteq\Bbb{Z}_m$, $p_m(A)|m$ and for each $a\in A$, if $t=|<a>|$, then we have $p_m(A)t=m$ and $t|\gcd(m,|A|)$. Moreover, $A=<a>$, for some $a\in A$, if and only if $p_m(A)|A|=m$.\\
\pf If $p_m(A)\not| m$, then there are some $q\in\Bbb{Z}$ and $0<r<p_m(A)$ such that $m=p_m(A)q+r$. Now, we have  $A+r=A-qp_m(A)=A$, since $A+p_m(A)=A$, which is a contradiction with definition of $p_m(A)$. Now, $t$ is the smallest positive number with $tp_m(A)=0\bmod m$, which implies that $tp_m(A)=m$, because $p_m(A)|m$. Similarly, for $t=|<a>|$, if $t\not|m$, then $m=tq+r$, for some $q\in\Bbb{Z}$ and $0<r\le t$. Now, $rp_m(A)=(m-tq)p_m(A)=0\bmod m$ which is a contradiction with definition of $t$, as the smallest positive integer satisfying $tp_m(A)=0\bmod m$. So, $t|m$. On the other hand, $|A|=\sum_{i=1}^k|<a_i>|=\sum_{i=1}^kt=kt$, so $t\mid |A|$, then, we have $t|\gcd(m,|A|)$. On the other hand, $A=<a>$, for some $a\in A$, if and only if $|A|=|<a>|=t$, or $|A|p_m(A)=m$. \eep
\label{rem1}
\end{rem}
\begin{rem}\rm
If $m$ is prime and $A\subseteq\Bbb{Z}_m$, then by Remark~\ref{rem1}, we have $p_m(A)|m$, so $p_m(A)=1$ or $p_m(A)=m$. However, $p_m(A)=1$ if and only if $A=\Bbb{Z}_m$, in this case $A=<a>$, for each $a \in A$.

\pf  If $p_m(A)=1$, then $t=m$, because $tp_m(A)=m$. So in this case, for each $a \in A$, $<a>=\{a+1\bmod m,a+2\bmod m,\ldots,a+m\bmod m\}=\Bbb{Z}_m$. \eep
\end{rem}
\begin{defn}\rm
Let $L=(i_0,j_0,i_1,j_1,\ldots,i_{l-1},j_{l-1})$ be a $2l$-cycle chain. If $e$, $1\le e\le l-1$, is the first number such that $(i_{e+p},j_{e+p})=(i_p,j_p)$, for all $0\le p\le l-1$, where all indices $e+p$ are reduced in modulo of $l$, then define $I(L)=\{(i_k,j_k),(i_{k+1},j_k):0\le k\le e-1\}$. Otherwise, if no such $e$ exists, then define $I(L)$ to be the empty set. In fact, $I(L)$ is the set of all pairs $(i_k,j_k)$, $1\le k\le l-1$ such that the cycle chain $L$ which starts from $(i_0,j_0)$ traverses the same points if it starts from $(i_k,j_k)$. Clearly, if $I(L)\ne\emptyset$, then $e|l$, so $n(L):=l/e$ is defined as the number of occurrence of $I(L)$ in the chain $L$, otherwise, we set $n(L)=1$. In fact, each $2l-$cycle ${\cal C}=v_1v_2\ldots v_{2l+1}$ ($v_{2l+1}=v_1$) in the Tanner graph (of the QC LDPC code with the parity-check matrix $H$) corresponding to the $2l-$cycle chain $L$, can be decomposed to $n(L)$ distinct paths ${\cal P}_k=v_{2e(k-1)+1}v_{2e(k-1)+2}\ldots v_{2ek+1}$, $1\le k\le n(L)$, such that for each $1\le i\le 2e$, all of the vertices $v_{i}$, $v_{2e+i}$, $\ldots$, $v_{2(n(L)-1)e+i}$ are in the same block of $H$.
\end{defn}
\begin{ex}\rm
\label{ex1}
Let $H$ be the parity-check matrix of a QC LDPC code shown in Fig.~\ref{fig1} with $(6,{\cal B})$-slope vector $(s_{0,0},s_{1,0},s_{0,1},s_{1,1},s_{0,2},s_{1,2})=(0,5,0,5,0,2),$ where ${\cal B}=\{\{1,2\},\{1,2\},\{1,2\}\}$. The dash lines in parts $(a)$ and $(b)$ indicate two $8-$cycle chains $L_1=(1,2,0,1,1,2,0,1)$ and $L_2=(1,2,0,0,1,2,0,1)$, respectively. It can be seen easily that $I(L_1)=\{(1,2),(0,2),(0,1),(1,1)\}$ and $I(L_2)=\emptyset$. For the chain $L_1$, we have $n(L_1)=2$, because $L_1$ includes two copy of the subchain $C=(1,2,0,1)$, i.e $L_1=(C|C)$, moreover, $n(L_2)=1$.
\end{ex}
We are now thus led to a theorem which gives the number of all distinct cycles in the Tanner graph of a QC-LDPC code corresponding to a given cycle-chain. First, we have two following lemmas.
\begin{lem}\rm
Let $m,s$ be two positive integers such that $s\le m$. Then, $\Bbb{Z}_m$, under the equivalence relation $a\sim b\Leftrightarrow s|b-a$,  can be partitioned to $a$ disjoint classes $[0]$, $[1]$, $\ldots$, $[a-1]$, $a=\gcd(s,m)$, each class has $m/a$ elements.

\pf Clearly, the equivalence classes from the partition of integer ring under the given relation are $[0]$, $[1]$, $\ldots$, $[s-1]$. On the other hand, in ring of the integers modulo of $m$, for each $i$, $0\le i\le a-1$, $a=\gcd(s,m)$, we have $[i]=[a+i]=[2a+i]=\ldots$, because $a=\alpha s+\beta m$, for some integers $\alpha$, $\beta$, so $a+i=\alpha s+\beta m+i=\alpha s+i$ in $\Bbb{Z}_m$. Hence, $a$ is divisible by $s$ in $\Bbb{Z}_m$ and so $[i]=[a+i]$.  Therefore, all of the disjoint classes in $\Bbb{Z}_m$ are $[0]$, $[1]$, $\ldots$, $[a-1]$.\eep
\label{lem2}
\end{lem}
\begin{lem}\rm
For each $2l$-cycle chain $L$, we have $n(L)|m$ and if $I(L)\ne\emptyset$, for each $(i,j)\in I(L)$, the set $A(i,j)$ has the period $m/n(L)$.

\pf
Let $L=(i_0,j_0,\cdots)$ be $n(L)$ copy of the subchain $C=(i_0,j_0,\cdots)$, i.e. $L=(C|C|\ldots|C)$ and ${\cal C}$ be the cycle in the Tanner graph  corresponding to the chain $L$. Moreover, let ${\cal P}_k$, $1\le k\le n(L)$, be the segment of $\cal C$ corresponding to the $k$th copy $C$ of $L$ which is a path in $\cal C$ with the starting and ending points in the $(i_0,j_0)$th block. It can be seen easily that the row-index difference of the end points in each path ${\cal P}_k$ is fixed, say the value $s\in\Bbb{Z}_m$. However, $\bigcup_{k=1}^{n(L)}{\cal P}_k$ is the cycle ${\cal C}$ with the same endpoints, so $n(L)s=0\pmod m$. On the other hand, $s\ne 0$ and $n(L)$ is the smallest number satisfying in $n(L)s=0\pmod m$, so $n(L)s=m$ which indicates $n(L)|m$. Now, for each $(i,j)\in I(L)$ and each path ${\cal P}_k$, $1\le k\le n(L)$, let $\{r^{(i,j)}_1,\cdots,r^{(i,j)}_l\}$ be the set of the row-indices of the points in the $(i,j)$th block. Clearly, $A(i,j)=\bigcup_{k=1}^l [r_k^{(i,j)}]$, where $[r_k^{(i,j)}]=\{r_k^{(i,j)}+ts\bmod m:0\le t<n(L)\}$, because among the points of $\cal C$ belong to the $(i,j)$th block, $s$ is the row-index difference of each point in ${\cal C}_k$ with the corresponding point in ${\cal C}_{k+1}$, for each $1\le k\le n(L)$ (clearly, the amount of $s$ is independent from $k$ and the selected points of ${\cal C}_k$). Now, the period of each $[r_k^{(i,j)}]$ is $s$, so $A(i,j)=A(i,j)+s\bmod m$. However, by the first part of the proof, we have $n(L)s=m$, so $s=m/n(L)$. Now, if $p=p_m(A(i,j))<s$, then $p$ is divisible by $s$, and the chain composing of $p$ subchain $C$ is a $2k-$cycle chain in $L$, for some $k<l$, which is a contradiction. \eep
\label{lem3}
\end{lem}
\begin{ex}\rm
For $L=(1,2,0,0,1,1,0,0,1,2,0,0,1,1,0,0)$ in Fig~.\ref{f2}, we have $I(L)=\{(i,j):0\le i\le 1, 0\le j\le 2\}$ and $n(L)=2$ which is divisible by $m=10$. On the other hand, $A(0,0)=A(1,0)=\{0,4,5,9\}$, $A(0,1)=A(1,2)=\{4,9\}$ and $A(0,2)=A(1,1)=\{0,5\}$, so $p_{10}(A(i,j))=5=m/n(L)$, for each $(i,j)\in I(L)$.
\end{ex}
\begin{thm}\rm
Each allowable $2l$-cycle chain $L=(i_0,j_0,i_1,j_1,\ldots,i_{l-1},j_{l-1})$ corresponds to $r=r(L)$ cycles of length $2l$ in the Tanner graph, where $r=m/n(L)$.

\pf Let ${\cal C}_1,{\cal C}_2,\ldots,{\cal C}_r$ be the all different $2l$-cycles in the Tanner graph corresponding to the cycle chain $L$. Moreover, let $L=(C|C|\ldots|C)$ be $n(L)$ copy of the subchain $C$ and for each $1\le t\le r$ and $1\le k\le n(L)$, let path ${\cal P}_k^{(t)}=v^{(t)}_{2(k-1)e+1}v^{(t)}_{2(k-1)e+2}\ldots v^{(t)}_{2ke+1}$, $e=l/n(L)$, be the segment of ${\cal C}_t=v^{(t)}_{1}v^{(t)}_{2}\ldots v^{(t)}_{2l}v^{(t)}_{2l+1}$, $v^{(t)}_1=v^{(t)}_{2l+1}$, corresponding to the $k$'th copy $C$ of $L$. Now, providing that $I(L)\ne\emptyset$, for each point $(i,j)\in I(L)=\{(i_k,j_k),(i_{k+1},j_k):0\le k\le l-1\}$, let $v^{(t)}_f$, $1\le t\le r$, be the first point of ${\cal P}_1^{(t)}$ belong to the $(i,j)$'th block of $H$ and $R^{(t)}(i,j)$ be the row indices of the points $v^{(t)}_f$, $v^{(t)}_{f+2e}$, $\ldots$, $v^{(t)}_{f+2(n(L)-1)e}$. Now, for each $1\le t\le r$, $p_m(R^{(t)}(i,j))=m/n(L)$, for each $1\le t\le r$. On the other hand, $\Bbb{Z}_m=\bigcup_{t=1}^rR^{(t)}(i,j)$, otherwise, there is a new $2l-$cycle corresponding to the chain $L$ starting from the point with the row index in $\Bbb{Z}_m\setminus\bigcup_{t=1}^rR^{(t)}(i,j)$. Hence, if we define $s:=p_m(R^{(t)}(i,j))=m/n(L)$, then $\Bbb{Z}_m$ can be partitioned to $r$ disjoint classes $R^{(t)}(i,j)$, $1\le t\le r$, therefore, by Lemma~\ref{lem2}, we have $r=\gcd(m,s)=m/n(L)$. On the other hand, $I(L)=\emptyset$, then each cycle ${\cal C}_t$, $1\le t\le r$, can be uniquely determined  from the starting point of the cycle in the block $(i_0,j_0)$ of $H$. Hence, if the starting points change, we have different cycles, so in this case $r=m=m/n(L)$ and the proof is completed.\eep
\label{thm1}
\end{thm}

\begin{figure}[h]
\begin{center}
\includegraphics[scale=0.8]{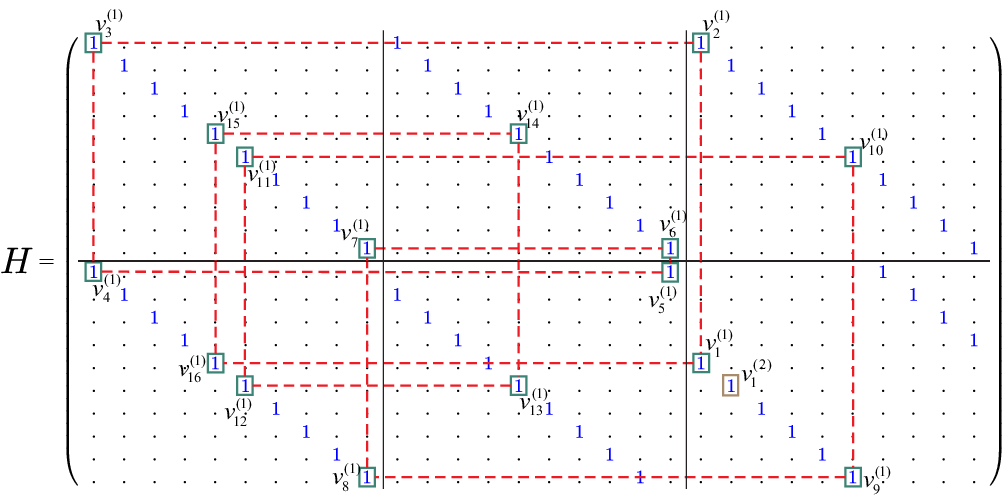}
\end{center}
\caption{\small The 16-cycle chain $(1,2,0,0,1,1,0,0,1,2,0,0,1,1,0,0)$ with $A(0,0)=A(1,0)=\{0,4,5,9\}$, $A(0,1)=A(1,2)=\{4,9\}$ and $A(0,2)=A(1,1)=\{0,5\}$ of period 5.} \label{nonb}
\end{figure}
To clarify the proof of Theorem~\ref{thm1}, we give the following example.
\begin{ex}\rm
In Fig.~\ref{nonb}, the 16-cycle ${\cal C}_1=v^{(1)}_1v^{(1)}_2\ldots v^{(1)}_{16}$ is given which corresponds to the 16-cycle chain $L=(1, 2, 0, 0, 1, 1, 0, 0, 1, 2, 0, 0, 1, 1, 0, 0)$. In fact, ${\cal C}_1$ can be partitioned to paths ${\cal P}^{(1)}_1=v^{(1)}_1v^{(1)}_2\ldots v^{(1)}_9$ and ${\cal P}^{(1)}_2=v^{(1)}_9v^{(1)}_{10}\ldots v^{(1)}_{16}v^{(1)}_1$. Then, $R^{(1)}(0,0)=R^{(1)}(1,0)=\{0,5\}$, $R^{(1)}(0,1)=\{9\}$, $R^{(1)}(1,2)=\{4\}$ and $R^{(1)}(0,2)=R^{(1)}(1,1)=\{0\}$. Similarly, for the cycle ${\cal C}_2$ starting from the $v_{1}^{(2)}$, we have $R^{(2)}(0,0)=R^{(2)}(1,0)=\{1,6\}$, $R^{(1)}(0,1)=\{0\}$, $R^{(1)}(1,2)=\{5\}$ and $R^{(1)}(0,2)=R^{(1)}(1,1)=\{1\}$. Counting this process,  $R^{(t)}(i,j)=R^{(1)}(i,j)+t\bmod 10$, for each $(i,j)\in I(L)$ and $1\le t\le5$, therefore, ${\Bbb Z}_{10}=\bigcup_{t=1}^5R^{(t)}(0,0)$.
\end{ex}
Now, the following result can be obtained obviously from Theorem~\ref{thm1}.
\begin{rem}\rm If $m$ is prime, then $r=1$ or $r=m$ (in Theorem~\ref{thm1}). In this case, if $m\ge l$, then $n(L)<m$, so $r=m$. Therefore, the number of $2l-$cycles in the QC LDPC code with CPM-size $m$ ($m$ is a prime not less than $l$), is $m$ times of the number of corresponding $2l-$cycle chains in the base matrix.
\end{rem}

Hereinafter, to simplify the notations, by the {\it cycle distribution} of an LDPC code with girth $2g$, we mean the polynomial $\lambda(x)=\sum_{i=g}^{\infty}\lambda_{2i}x^{2i}$, in which $\lambda_{2i}$ is the number of $2i$-cycles in the Tanner graph of the code.
\begin{ex}\rm\label{ex3}
Let $H$ be the following parity-check matrix corresponding to a (3,4) QC-LDPC code with girth 12 and CPM-size 100.
$$H=\left(
     \begin{array}{cccc}
       {\cal I} & {\cal I} & {\cal I} & {\cal I} \\
       {\cal I} & {\cal I}^1 & {\cal I}^7 & {\cal I}^{30} \\
       {\cal I} & {\cal I}^3 & {\cal I}^{19} & {\cal I}^{75} \\
     \end{array}
   \right)$$
  Using the algorithm given in next section, the cycle distribution of the code is $\lambda(x)=6000x^{12}+24400x^{14}+99825x^{16}+550500x^{18}+3052200x^{20}+12793400x^{22}+21587550x^{24}+\cdots$. As it can be seen from Theorem~\ref{thm1}, to find the cycle distribution, first we must find all of the allowable (non-isomorph) cycle chains. For example, all of the 12-cycle chains in $H$ are provided in Table~\ref{tab1}. Although, for each 12-cycle chain $L$, we have $r(L)=m=100$, this is not true in general for cycle chains with greater length. For example, we have 1000 allowable (non-isomorph) 16-cycle chains which just there of them have $r<m$, i.e.
  $L_1=(2,3,0,0,2,3,0,0,2,3,0,0,2,3,0,0)$, $L_2=(2,3,0,2,1,1,0,2,2,3,0,2,1,1,0,2)$ and $L_3=(2,3,1,2,0,0,1,1,2,3,1,2,0,0,1,1)$, with $r(L_1)=25$, $r(L_2)=50$ and $r(L_3)=50$.
\end{ex}
\begin{table}\scriptsize
\[\begin{array}{c|c|c}
\begin{array}{c|c}{\rm num}&12-{\rm cycle\,\,chain}\\\hline\hline
  1&(2,1,0,0,1,1,2,0,0,1,1,0)\\\hline
2&(2,1,1,0,0,1,1,0,0,1,1,0)\\\hline
3&(1,2,0,0,1,1,0,2,1,0,0,1)\\\hline
4&(1,2,0,1,2,0,0,1,2,0,0,1)\\\hline
5&(2,2,0,0,1,2,2,0,0,2,1,0)\\\hline
6&(2,2,0,0,2,1,0,2,2,0,0,1)\\\hline
7&(2,2,0,1,1,2,2,1,0,2,1,1)\\\hline
8&(2,2,1,0,0,1,2,0,0,2,1,1)\\\hline
9&(2,2,1,0,0,2,1,1,2,0,0,1)\\\hline
10&(2,2,1,0,2,1,1,2,2,0,1,1)\\\hline
11&(2,2,1,1,0,2,1,1,0,2,1,0)\\\hline
12&(2,2,1,1,2,0,0,2,1,0,0,1)\\\hline
13&(1,3,0,0,1,1,0,3,1,0,0,1)\\\hline
14&(1,3,0,0,1,2,0,3,1,0,0,2)\\\hline
15&(1,3,0,1,1,2,0,3,1,1,0,2)\\\hline
16&(1,3,0,1,1,2,2,0,0,2,2,1)\\\hline
17&(1,3,0,2,1,0,0,2,2,1,0,2)\\\hline
18&(1,3,0,2,2,0,0,1,1,2,2,1)\\\hline
19&(1,3,0,2,2,1,0,2,1,0,0,2)\\\hline
20&(1,3,0,2,2,1,1,2,2,0,0,1)\end{array}&
\begin{array}{c|c}{\rm num}&12-{\rm cycle\,\,chain}\\\hline\hline
21&(2,3,0,0,1,1,0,2,2,3,1,1)\\\hline
22&(2,3,0,0,1,1,2,0,1,3,0,1)\\\hline
23&(2,3,0,0,1,1,2,3,1,1,0,2)\\\hline
24&(2,3,0,0,1,3,0,1,2,0,1,1)\\\hline
25&(2,3,0,0,1,3,2,0,0,3,1,0)\\\hline
26&(2,3,0,0,2,1,0,3,2,0,0,1)\\\hline
27&(2,3,0,0,2,1,1,3,0,2,1,0)\\\hline
28&(2,3,0,0,2,2,0,1,1,2,0,0)\\\hline
29&(2,3,0,0,2,2,0,3,2,0,0,2)\\\hline
30&(2,3,0,1,1,0,0,1,1,3,0,1)\\\hline
31&(2,3,0,1,1,0,0,2,2,3,1,0)\\\hline
32&(2,3,0,1,1,0,2,3,1,0,0,2)\\\hline
33&(2,3,0,1,1,2,0,0,2,2,0,0)\\\hline
34&(2,3,0,1,1,3,0,1,1,0,0,1)\\\hline
35&(2,3,0,1,1,3,2,1,0,3,1,1)\\\hline
36&(2,3,0,1,2,0,1,3,0,0,1,1)\\\hline
37&(2,3,0,1,2,2,0,1,2,2,1,0)\\\hline
38&(2,3,0,1,2,2,0,3,2,1,0,2)\\\hline
39&(2,3,0,1,2,2,1,0,2,2,0,1)\\\hline
40&(2,3,0,1,2,3,0,1,2,3,0,2)\end{array}&
\begin{array}{c|c}{\rm num}&12-{\rm cycle\,\,chain}\\\hline\hline
41&(2,3,0,2,1,0,2,1,1,3,0,0)\\\hline
42&(2,3,0,2,1,3,0,0,2,1,1,0)\\\hline
43&(2,3,0,2,1,3,2,2,0,3,1,2)\\\hline
44&(2,3,0,2,2,0,1,2,2,3,0,2)\\\hline
45&(2,3,0,2,2,3,1,0,0,1,1,0)\\\hline
46&(2,3,0,2,2,3,1,1,0,0,1,1)\\\hline
47&(2,3,1,0,0,1,2,0,0,3,1,2)\\\hline
48&(2,3,1,0,0,3,1,2,2,0,0,1)\\\hline
49&(2,3,1,0,2,1,0,3,1,1,0,2)\\\hline
50&(2,3,1,0,2,1,1,3,2,0,1,1)\\\hline
51&(2,3,1,0,2,2,1,0,2,3,1,1)\\\hline
52&(2,3,1,0,2,2,1,1,2,3,1,0)\\\hline
53&(2,3,1,0,2,2,1,3,2,0,1,2)\\\hline
54&(2,3,1,0,2,3,1,1,2,2,1,0)\\\hline
55&(2,3,1,1,0,2,2,1,0,3,1,0)\\\hline
56&(2,3,1,1,0,3,1,0,2,1,0,2)\\\hline
57&(2,3,1,1,2,2,1,3,2,1,1,2)\\\hline
58&(2,3,1,2,0,3,1,2,2,1,1,2)\\\hline
59&(2,3,1,2,2,0,0,3,1,0,0,1)\\\hline
60&(2,3,1,2,2,1,1,2,0,3,1,2)
\end{array}\end{array}\]
\caption{\label{tab1}All of the (non-isomorph) allowable 12-cycle chains in Example~\ref{ex3}}
\end{table}

In continue, using Theorem~\ref{thm1}, an algorithm is proposed which efficiently finds the cycles (with arbitrary lengths not less than the girth) in the Tanner graph of a QC-LDPC code by investigating the cycle chains. To do this, first we pursue the cycle chains in the parity-check matrix column by column, from the top to the bottom, then Theorem~\ref{thm1} is used to find the number of cycles corresponding to each cycle chain.

\section{An Efficient Algorithm for Counting the Cycles}
\label{sec4}
For given positive integers $b$, $v$ and $k$, let $B$ be a $v\times k$ binary matrix with the corresponding $(v,k)-$design ${\cal B}=[B_1,\ldots,B_k]$. Moreover, let $m\ge 1$ be an integer and $S$ be a $(m,{\cal B})-$slope vector such that the girth of the QC LDPC code with the parity-check matrix ${\cal H}_{m,{\cal B},S}$ is $2g$.
Here, we propose a deterministic algorithm to enumerate all of the cycles in TG$({\cal H}_{m,{\cal B},S})$ up to $2l$ in which $l$ is an arbitrary positive integer not less than $g$. It is noticed that, unlike the known counting algorithms which count the cycles up to length at most $2g-2$, the proposed algorithm is capable to count cycles of length $2l$, for each $l\ge 2$. In the algorithm, to classify non-isomorphic cycle chains and in order to speed up the process, for a given cycle chain $L=(i_0,j_0,i_1,j_1,\cdots,i_{l-1},j_{l-1})$, we use the functions $b^{\to}(L)$ and $b^{\leftarrow}(L)$ to be the $k$-adic representation of $L$ to the right and left, respectively, i.e. $b^{\to}(L)=\sum_{t=0}^{l-1}(i_t+j_tk)k^{2t}=i_0+j_0k+i_1k^2+j_2k^3+\cdots+i_{l-1}k^{2l-2}+j_{l-1}k^{2l-1}$ and $b^{\leftarrow}(L)=\sum_{t=0}^{l-1}(i_{l-t}+j_{l-t-1}k)k^{2t}=i_0+j_{l-1}k+i_{l-1}k^2+j_{l-2}k^3+\cdots+j_0k^{2l-1}$. Now, it can be seen easily that two $2l$-chains $L_1$ and $L_2$ are isomorphic if and only if $b^\to(L_1)=k^{2t}b^\to(L_2)\pmod{k^{2l}-1}$ or  $b^\to(L_1)=k^{2t}b^{\leftarrow}(L_2)\pmod{k^{2l}-1}$, for some $0\le t\le l-1$. In the algorithm, by ${\cal B}^{(e)}$, $1\le e\le \sum_{i=1}^k(|B_i|-1)$, we mean the first $e$ elements of $\cal B$, when the elements in the blocks (except the first element in each block) are enumerated one by one from the left to the right. On the other hand, ${\cal B}^{(e)}=[B_1,\cdots,B_{p-1},B'_{p}]$, where $p$ is the largest positive integer satisfying in $\sum_{i=1}^{p-1}(|B_i|-1)<e$, and $B'_{p}$ is the first $e+1-\sum_{i=1}^{p-1}(|B_i|-1)$ elements of $B_p$. For example, if ${\cal B}=[\{1,2,3\},\{3,4,5,6\}]$, then ${\cal B}^{(1)}=[\{1,2\}]$, ${\cal B}^{(2)}=[\{1,2,3\}]$, ${\cal B}^{(3)}=[\{1,2,3\},\{3,4\}]$, ${\cal B}^{(4)}=[\{1,2,3\},\{3,4,5\}]$ and  ${\cal B}^{(5)}=[\{1,2,3\},\{3,4,5,6\}]={\cal B}$. Now, the outline of the algorithm is as follows.

\newpage
\begin{alg}
\begin{algorithmic}
\STATE{$n\leftarrow\sum_{i=1}^{k}(|B_i|-1)$ and ${\cal A}\leftarrow\emptyset$}
\FOR{$e$ from 1 \TO $n$}
 \STATE{Let ${\cal B}^{(e)}=[B_1,\cdots,B_{p-1},B'_p]$, where $B'_{p}=\{b_0,b_1,\cdots,b_{u}\}\subseteq B_p$, for $u=e-\sum_{i=1}^{p-1}(|B_i|-1)$.}
 \FOR{$i$ from $0$ \TO $u$}
  \FOR{$j$ from $i+1$ \TO $u$}
   \STATE{Let $L=(b_i,p,b_j,\cdots)$ be a $2l-$alowable cycle chain in ${\cal B}^{(e)}$ starting from $(b_i,p,b_j)$.}
   \IF{$L$ is not isomorphic with elements of $\cal A$}
    \STATE{${\cal A}\leftarrow{\cal A}\cup\{L\}$.}
   \ENDIF
  \ENDFOR
 \ENDFOR
\ENDFOR
\RETURN{$\sum\limits_{L\in{\cal A}}r(L)$ as the number of $2l$ cycles.}
\end{algorithmic}
\label{alg1}
\end{alg}
In fact, Algorithm~\ref{alg1} counts the cycles of length $2l$ by $2l-$cycle chains sequentially from the first $e$, $1\le e\le n$, elements of $\cal B$, denoted by ${\cal B}^{(e)}$. For this, first, Lemma~\ref{lem-allow-qc} is used to investigate allowability of the constructed cycle chain. Then, each of the previousely constructed allowable $2l-$cycle chains is compared with $L$ to be non-isomorphic. This process is the only part of the algorithm which needs some times for running, causing a complexity.
For this problem, as mentioned above, $b^{\to}(L)$ and $b^{\leftarrow}(L)$ can be useful to speed up this process of the algorithm. Finally, Theorem~\ref{thm1} is used to find $2l-$cycles corresponding to each $2l-$cycle chain. It is noticed that in $e$th step of the algorithm, to find the allowable cycle chains and corresponding $2l$ cycles, we just consider the parity-check matrix ${\cal H}^{(e)}$, which is a submatrix of $\cal H$ constructed in step $e$ based on the design ${\cal B}^{(e)}$.


\begin{ex}\rm
In this example, we use Algorithm~\ref{alg1} to count the short cycles in the Tanner graph of some standard QC-LDPC codes. The outputs were obtained by a C$\sharp$ programming applied on a computer with a 2.2-GHz CPU and 6 GB of RAM. Consider two LDPC codes adopted in IEEE 802.11 standard \cite{istandard} with rate-2/3. These codes are two irregular (1296, 432) and (1944,648) QC-LDPC codes denoted by $A$ and $B$, respectively. Applying Algorithm~\ref{alg1}, Table \ref{mytab1} provides the multiplicity of $2l-$cycles, $2\le l\le 6$, denoted by $N_{2l}$. Moreover, the running time is compared with the time of a counting algorithm in \cite{bani}. While the algorithm in \cite{bani} can only compute $N_4$, $N_6$ and $N_8$, the proposed algorithm can enumerate ${2l}-$cycle multiplicities, for each $l\ge 2$.
Moreover, the running time of Algorithm~\ref{alg1} is remarkably less than the time of Algorithm in \cite{bani}. In fact, for codes $A$ and $B$, the running times of Algorithm~\ref{alg1} are 0.046 and 0.048 seconds, respectively, while the consumed time in \cite{bani} are about 0.5 and 1 seconds, respectively.
\label{ex4}
\end{ex}
\begin{table}
\begin{center}
\begin{tabular}{|l|l||l|l|l|l|l|}
\hline
& &\multicolumn{5}{c|}{Number  of Cycles}\\
\cline{3-7}
Code &  Length &$N_4$&$N_6$&$N_8$&$N_{10}$&$N_{12}$\\
\hline\hline
$A$&1296&108&7830&237627&6884028&198486018\\
\hline
$B$&1944&81&6399&251667&7071624&211628106\\
\hline
\end{tabular}
\caption{\label{mytab1} Cycle multiplicities of codes $A$ and $B$ in Example~\ref{ex4}.}
\end{center}
\end{table}

\subsection{The Complexity of The Algorithm}
There is a main problem to verify the complexity of Algorithm~\ref{alg1} as the main conclusion of the paper. First, let $L=\sum_{i=1}^{k}(|B_i|-1)$ be the number of steps which must be passed to reach the solution. In fact, $L$ is the number of 1's in the base matrix except the first ones in the columns when traversing the base matrix from up to down. In step $e$, we must find all non-isomorphic chains starting from $(b_i,p,b_j)$, inquires checking all (allowable) chains $(i_0,j_0,i_1,j_1,\cdots,i_{l-1},j_{l-1})$ with $(i_l,j_l)=(i_0,j_0)=(b_i,p)$, $i_1=b_j$ and $\{i_k,i_{k+1}\}\subseteq B_{j_k}$, $1\le k\le l-1$. In this case, if $b_{\max}=\max_{1\le i\le k}|B_i|$, then such chains can be examined in at most $(b_{\max}-1)^{l-2}(p-1)^{l-1}$ which are compared with at most $2l\sum_{p'=1}^p(b_{\max}-1)^{l-2}(p'-1)^{l-1}$ chains to verify the non-isomorphic relation. Thus, the overall complexity to find all non-isomorphic chains are at most $2l\sum_{p=1}^k(b_{\max}-1)^{l-2}(p-1)^{l-1}\sum_{p'=1}^p(b_{\max}-1)^{l-2}(p'-1)^{l-1}=2l\sum_{p=1}^k\sum_{p'=1}^p(b_{\max}-1)^{2l-4}(p-1)^{l-1}(p'-1)^{l-1}\in O(lb_{\max}^{2l}k^{2l})$ which is polynomial by $k$, i.e. the number of blocks, if $l$ and $b_{\max}$ are given.
\section{Conclusion}
In this paper, an efficient algorithm for counting short cycles in the Tanner graph of a QC-LDPC code is presented. Although, the known counting algorithms can enumerate cycles up to $2g-2$, where $g$ is the girth of the code, the proposed algorithm is capable of counting any (even) cycles of length at least $g$ in the Tanner graph. Interestingly, for QC-LDPC codes lifted from a protograph $G$, the complexity of the algorithm is based on the number of edges in $G$, independent from the lifting degree of the constructed codes. Finally, applying the proposed algorithm for some standard codes, the overall complexity improves rather than the known algorithms, in terms of computational complexity and memory requirements.
\section{Acknowledgments}
We would like to thank the anonymous referees for their helpful comments. This work was supported in part by the research council of Shahrekord university. 

\end{document}